 \documentclass[final,5p,times,twocolumn]{elsarticle}

\usepackage[colorlinks]{hyperref}
\hypersetup{linkcolor=blue,citecolor=blue,urlcolor=blue}
\usepackage{dcolumn}
\usepackage{braket}
\usepackage{amssymb}
\usepackage{amsmath}
\usepackage{lineno}

\journal{Physics Letters B}

\begin{document}
\begin{frontmatter}



\title{Role of two-body dissipation on the mean-field dynamics validity}


\author[first]{Yingge Huang}
\author[first]{Hui Wang}
\author[first]{Erxi Xiao}
\author[first]{Long Zhu}
\author[first,second]{Jun Su\corref{mycorrespondingauthor}}
\cortext[mycorrespondingauthor]{Corresponding author: sujun3@mail.sysu.edu.cn}

\affiliation[first]{organization={Sino-French Institute of Nuclear Engineering and Technology},
            addressline={Sun Yat-sen University}, 
            city={Zhuhai},
            postcode={519082}, 
            country={China}}
\affiliation[second]{organization={Key Laboratory of Nuclear Data},
            addressline={China Institute of Atomic Energy}, 
            city={Beijing},
            postcode={102413}, 
            country={China}}

\begin{abstract}

The role of two-body dissipation in nuclear reactions at energies of several times Coulomb barrier remains unclear but is crucial for understanding the mechanisms of deep-inelastic reactions.
In this letter, we report a systematic analysis of two-body dissipation effects on the validity of mean-field dynamics, enabled by the TDHF-QRx approach, which incorporates the collision term via the relaxation-time approximation rather than full collision calculations.
For deep-inelastic reactions, the contact time between nuclei is found to increase, resulting in changes to the reaction process and fragment properties such as scattering angles and total kinetic energy.
These changes become important with the increase of reaction energy and decrease of impact parameter.
We identify the range of reaction condition where two-body dissipation becomes significant, providing valuable insights for the applicability of mean-field dynamics approaches.
The limitations of model, particularly those arising from the incomplete conservation of the locality of two-body dissipation within the quantum framework, are also discussed.

\end{abstract}



\begin{keyword}
two-body dissipation \sep relaxation-time approximation \sep TDHF-QRx \sep heavy-ion collisions



\end{keyword}

\end{frontmatter}




\section{Introduction}
\label{introduction}

Energy dissipation, fundamental to non-equilibrium systems like nuclear reactions \cite{PhysRevC.79.054606,PhysRevC.103.034603,PhysRevC.97.014603,PhysRevC.90.054605,PhysRevC.110.054605,PhysRevC.27.590,PhysRevC.88.044611}, is categorized into one-body, two-body, and many-body mechanisms based on the Bogoliubov-Born-Green-Kirkwood-Yvon (BBGKY) hierarchy. 
One-body dissipation refers to the loss of collective energy driven by single-particle evolution in the mean field.
Two-body dissipation, arising from nucleon pair interactions, becomes significant at intermediate energies, typically several times the Coulomb barrier. 
This energy range features interplay between one- and two-body dissipation,
and also complex reaction mechanisms such as fusion, deep-inelastic scattering, and multi-fragmentation, posing a key challenge for quantum many-body dynamics.

The time-dependent Hartree-Fock approach (TDHF), one of the primary microscopic mean-field approach, has achieved remarkable success in describing low-energy nuclear dynamics \cite{SIMENEL201819,Simenel2012}, particularly in capturing reaction dissipation through its one-body dissipation mechanism \cite{PhysRevLett.120.022501}.
The rationale for neglecting two-body dissipation at low energies is rooted in the Pauli exclusion principle, which inhibits nucleon-nucleon (NN) collisions. 
In the intermediate energy region, two-body dissipation, usually described by the semiclassical NN cascade collision method, is found to play a critical role in modeling spallation and fragmentation reactions \cite{PhysRevC.97.034625,ONO2004501}.
It suggests the single-Slater-determinant wavefunction in TDHF becomes inadequate for describing the complex excited configurations generated by NN collisions.
However, it remains unclear under which conditions two-body dissipation becomes significant and how it influences the mean-field dynamics \cite{SIMENEL201819,Dinh2018}, which is also related to understanding the nuclear equation of state \cite{PhysRevC.111.014609}. 
Several approaches have been proposed to address this issue within quantum framework.
The time-dependent density matrix (TDDM) \cite{10.3389/fphy.2020.00067} and TDDM$^P$ \cite{PhysRevLett.102.202501} methods directly solve the two-body correlations, including pairing and NN collision effects.
Due to the huge computational resource requirement for fully considering the two-body degree of freedom, and the problem of zero range Skyrme interactions, TDDM is not yet widely applicable on heavy-ion collisions (HICs) \cite{lacroix2014stochastic,PhysRevC.65.037601,PhysRevC.93.034607,PhysRevC.98.014603,PhysRevC.103.064304}. 
The approaches including only the NN collision correlations, such as Extended-TDHF \cite{LACROIX2004497} and stochastic TDHF \cite{lacroix2014stochastic}, provide part of insights into NN collision effects in the nuclear reactions. 
Addressing the role of two-body dissipation remains a challenge due to the computational complexity associated with the NN collisions.

A practical approach is to bypass the full treatment of NN collisions and instead directly incorporate the momentum thermalization effects, leading to the concept of the relaxation-time approximation (RTA).
The RTA was first introduced in nuclear HICs in the 1980s, with phenomenological treatment of the equilibrium momentum distributions \cite{KOHLER1980315,KOHLER1988318}. 
A decade ago, the quantum RTA (QRTA) framework was proposed for fermion systems, enabling equilibrium-state calculations in a fully quantum-mechanical manner \cite{REINHARD2015183}. 
While this framework has been successfully applied in atomic and molecular systems \cite{Vincendon2017,Hughes2023}, its application to nuclear systems remains limited \cite{PhysRevC.111.044604}.

In this work, the QRTA framework is employed to incorporate two-body dissipation mechanisms into the Skyrme-TDHF approach, referred to as the TDHF-QRx (\textbf{Q}uantum \textbf{R}ela\textbf{x}ation) following the name of the pioneering work \cite{PhysRevC.40.1711}. 
The binary HICs at energies above the fusion barrier are computed using both TDHF and TDHF-QRx, to reveal the impact of the two-body dissipations to the reaction dynamics and fragment properties.
Through systematic calculations, we determine the critical effective range where two-body dissipation should be considered for describing reaction dynamics.
The limitations of the current QRTA framework are also discussed, with particular attention to those stemming from the simplified treatment of the locality in two-body dissipation.

\section{\label{sec:theory}Theoretical framework and implementation}

To describe the nuclear dynamics within a fully quantum framework, the starting theory is the TDHF under the mean-field approximation, which is expressed as
 \begin{equation}
	i\hbar\frac{\partial \hat \rho_{1}}{\partial t} = \left [  \hat h_1, \hat \rho_{1}\right ]. 
 \end{equation}
where $\hat \rho_{1}=\sum_k W_k \ket{\varphi_k}  \bra{\varphi_k}$ represents the one-body density operator with the single-particle (sp) states $\varphi_k$ and the occupations $W_k$,
and the $\hat h_1$ is the sp hamiltonian.
The two-body correlations can be incorporated based on the BBGKY hierarchy equations truncated at the first order \cite{bonitz2016quantum}
 \begin{equation}\label{eq1}
	i\hbar\frac{\partial \hat \rho_{1}}{\partial t} =  \left [  \hat t_1, \hat \rho_{1} \right] + \mathrm{Tr}_{2}(\left [  \hat V_{12}, \hat \rho_{12}\right ]).
 \end{equation}
Equation~\eqref{eq1} can be simplified by considering only the correlations arising from NN collisions, as pairing correlations are less significant in systems with high excitation energies \cite{lacroix2014stochastic}.
It leads to the quantum Boltzmann equation
\begin{equation}\label{eq3}
i\hbar\frac{\partial{\hat \rho}}{\partial t} = \left[ \hat h[\hat \rho], \hat \rho \right] + \hat I_{\mathrm{coll}}[\hat \rho],
\end{equation}
where $\hat I_{\mathrm{coll}}$ is the collision term,
and the index 1 is omitted here and below for simplifying the notation.
The collision term in the semi-classical Boltzmann transport equation can be approximated using a linear term when system is close to equilibrium, known as the relaxation-time approximation (RTA). 
Motivated by the RTA, the Eq.~\eqref{eq3} can be rewritten by applying quantum RTA (QRTA) 
\begin{equation}\label{eq4}
\frac{\partial{\hat \rho}}{\partial t} = - \frac{i}{\hbar}[\hat h, \hat \rho]- \frac{1}{\tau_{\mathrm{relax}}}\left(\hat{\rho} - \hat{\rho}_{\mathrm{eq}}[\rho,\boldsymbol{j},E_{\mathrm{tot}}] \right),
\end{equation}
where $\tau_{\mathrm{relax}}$ is the relaxation time, and the $\hat{\rho}_{\mathrm{eq}}$ is the density operator related to the equilibrium state.
To broadly evaluate the effects of two-body dissipation, we assume an extended application of the linear form of NN collisions in nuclear reactions.
In principle, the NN collisions should be considered locally.
Accordingly, in the semi-classical RTA, the density $\rho$, the current $\boldsymbol{j}$, and the kinetic-energy density are conserved at each spatial point.
However, this approach cannot be completely applied in QRTA.
Since the semi-classical local kinetic-energy density concept is ambiguous in quantum systems, the total energy $E_{\mathrm{tot}}$ of the whole system is instead used as a conserved quantity in the equilibrium state calculations \cite{REINHARD2015183}.


The code of TDHF-QRx approach is developed based on the TDHF code Sky3D \cite{ABHISHEK2024109239}.
The strategy of the QRTA calculation refers to the QDD code of atomic system \cite{DINH2022108155}.
Before the dynamics calculations, the ground-state of the projectile and the target nucleus are computed by the static Hatree-Fock (HF). 
To eliminate center-of-mass corrections, the Skyrme interaction parameterization utilized is SLy4d \cite{Kim_1997}.
The dynamic calculations are performed in a $(60\times60\times24) \Delta d^3$ cartesian mesh with $\Delta d=0.8~\rm{fm}$, and the time step $\Delta t=0.2$ fm/$c$.
At initial moment, they are positioned 17.6 fm apart.
The QRTA calculation can perform in a larger time step than $\Delta t$ by a factor of 50-200, as nucleons move more slowly due to relaxation compared to mean-field propagation \cite{DINH2022108155}.
In present work, $\Delta t_{\rm{RTA}}$ is set to 100$\Delta t$, determined by checking the sensitivity to the evolution of collective energies. 
For the steps without computing the equilibrium states, the propagation of the states is similar with TDHF.
The details of the TDHF can be found in Ref.~\cite{ABHISHEK2024109239,simenel2010quantum}

When the propagation reaches the QRTA step (e.g., $t_{n}$), the processes of the QRTA calculations are the following.
First is computing the equilibrium states.
We note $\hat{\rho}_{\mathrm{mf}}$ the states just after the evolution from $t_{n-1}$ step to $t_{n}$.
After computing the density $\rho$, probability current $\boldsymbol{j}$, and total energy at $t_{n}$,
a static HF calculation is performed to calculate the equilibrium states with density constraint and current constraint. 
References \cite{PhysRevC.32.172,cusson1985density} give details of the constraint HF calculations. 
During the HF iterations, the occupations $W_k$ of equilibrium states are needed to adjust to the Fermi distribution 
\begin{equation}\label{eq:fermi}
W_k^{\left( \mathrm{eq} \right)}=\frac{1}{1+\exp \left( \left( \epsilon _k-\mu \right) /k_{\mathrm{B}}T^\prime \right)} ,
\end{equation}
where $\epsilon_k$ is the sp energy, $k_{\mathrm{B}}$ is the Boltzmann constant, and $\mu$ and $T^\prime$ are adjustable parameters representing the chemical potential and temperature, respectively.
These varying occupations allow for the conservation of total energy $E_{\mathrm{tot}}$ during the equilibrium state calculation.
When the iterations converge, we obtain the $\hat{\rho}_{\mathrm{eq}}$ for the equilibrium states with the adjusted occupations, and the intrinsic excitation energy $E^*_{\mathrm{intr}}$. 
$E^*_{\mathrm{intr}}$ is calculated as
$E^*_{\mathrm{intr}}=E_{\mathrm{tot}}-E_{\mathrm{CHF}}\left[ \rho ,\boldsymbol{j},T^\prime=0 \right]$,
where $E_{\mathrm{CHF}}\left[ \rho ,\boldsymbol{j},T^\prime=0 \right]$  is evaluated by adjusting the occupations of the constrained state to the zero-temperature distribution.

Secondly, we mix the $\hat \rho_\mathrm{eq}$ into the $\hat{\rho}_{\mathrm{mf}}$ to incorporate the two-body dissipation effects.
The relaxation time as the mixing factor, evaluated from the homogeneous nuclear matter, is computed as \cite{bertsch1978collision}
\begin{equation}\label{eq:rt}
	\frac{\hbar}{\tau_{\mathrm{relax}}}=7.9 \frac{\hbar^{2}}{m} \sigma_{NN} k_{\mathrm{F}} \rho_{0} \frac{T^{2}}{\varepsilon_{\mathrm{F}}^{2}},
\end{equation}
where $m$ is the nucleon mass, $k_{\mathrm{F}}$ is the Fermi momentum, $\rho_0$ is the saturation density of nuclear matter, and $\varepsilon_{\mathrm{F}}$ is the Fermi energy. 
The effective in-medium NN collision cross-section, $\sigma_{NN}$, is set to the typical value of 40 mb in this letter, unless explicitly stated otherwise.
From a physical perspective, the relaxation time should be evaluated using the local density and temperature. 
Since the implementation of the $\boldsymbol{r}$-dependent relaxation time is cumbersome within quantum mechanical framework,
the global saturation density $\rho_0$ and temperature $T$ are chosen as approximate values in Eq.~\ref{eq:rt}. 
$T$ is calculated by
\begin{equation}\label{eq:tem}
\frac{\pi ^2T^2}{4\varepsilon _{\mathrm{F}}}=\frac{E_{\mathrm{intr}}^{*}}{N} ,
\end{equation}
where $N$ is the particle number and $\varepsilon_{\rm F}$ is the Fermi energy.
The $\hat{\rho}_{\mathrm{mix}}$ incorporating the two-body dissipations is computed as
	\begin{equation}\label{eq:mix}
	 \begin{aligned}
\hat{\rho}_\mathrm{mix}\left(t_{n+1}\right) & = (1-\eta) \hat{\rho}_{\mathrm{mf}}+\eta \hat{\rho}_{\mathrm{eq}},\\
 \quad \eta & = \frac{\Delta t_{\mathrm{RTA}}}{\tau_{\mathrm{relax }}}.
	 \end{aligned}
	\end{equation}
Completing this addition requires transforming the base of the equilibrium state to the one of mean-field state.
Diagonalization of ${\rho}_{\mathrm{mix}}$ is then performed to obtain the set of states $\left\{\ket{\varphi_{k}^{n}}\right\}$ in natural-orbital representation.
And the next time step in the evolution can be proceed.

The TDHF and its extensions would not be valid for long time calculations. 
Based on the results in Ref.~\cite{SM_PRL120.022501} and considering computational resources, the maximum duration for a single-event calculation is set to 1600 fm/$c$ in this work.
An event is classified as a fusion reaction if the dinuclear system remains unseparated when this time limit is reached.

\section{Results and discussions}

\begin{figure}
\center
\includegraphics[width=8.2cm]{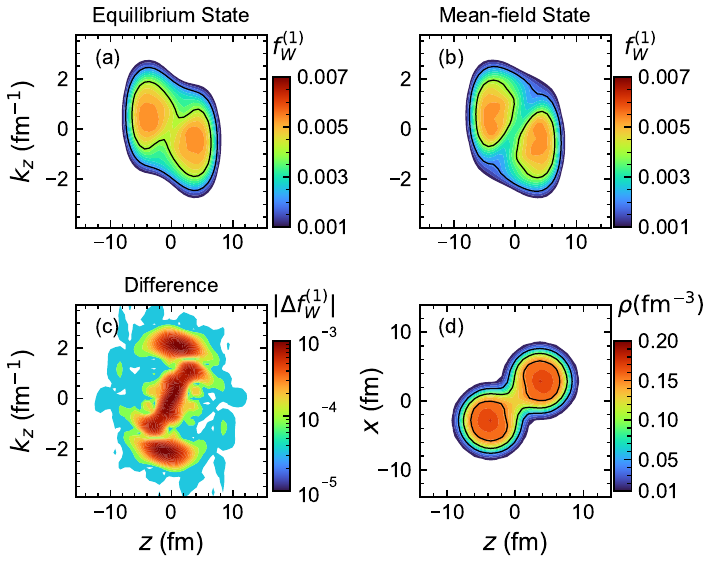} 
\caption{\label{fig:locality}(a)(b) The one-dimensional Wigner distributions of the equilibrium state $\hat{\rho}_{\mathrm{eq}}$ and the mean-field state $\hat{\rho}_{\mathrm{mf}}$, respectively, for the $^{60}\rm{Ni}+{^{60}\rm{Ni}}$ collision at $t=80~\rm{fm/}c$, $E_{\rm{c.m.}}=400~\rm{MeV}$, and impact parameter $b=7.5~\rm{fm}$. (c) The absolute difference between (a) and (b). (d) The corresponding density distribution.}
\end{figure}
The $^{60}\rm{Ni}+{^{60}\rm{Ni}}$ binary reaction at $E_{\rm{c.m.}}=400~\rm{MeV}$ ($\sim4.1V_{\rm B}$) with impact parameter $b=7.5~\rm{fm}$ is first calculated to illustrate the thermalization of momentum distribution induced by the two-body dissipation.  
This system is chosen regarding the exclusion of isospin effects, which could contribute to dissipation due to the close relationship between equilibrium and dissipation \cite{PhysRevLett.124.212504}, allowing a focused study on the two-body dissipation.
Additionally, dissipation in this system is demonstrated being well described at near-barrier energies by the mean-field approach \cite{PhysRevLett.120.022501}.
Following the formalism of Ref.~\cite{PhysRevC.84.034608}, Fig.~\ref{fig:locality}(a) presents the one-dimensional Wigner distribution of $\hat{\rho}_{\mathrm{eq}}$ at the early stage of reaction, which shows an enhancement of momentum thermalization comparing to the $\hat{\rho}_{\mathrm{mf}}$ (Fig.~\ref{fig:locality}(b)).
This thermalization mainly occurs in the overlap region between two nuclei, as shown in Fig.~\ref{fig:locality}(c), corresponding to the spatial domain where NN collisions take place.
However, due to the simplification in kinetic-energy constraint, a certain degree of loss of locality of two-body dissipation is present.

\begin{figure}
\center
\includegraphics[width=7.2cm]{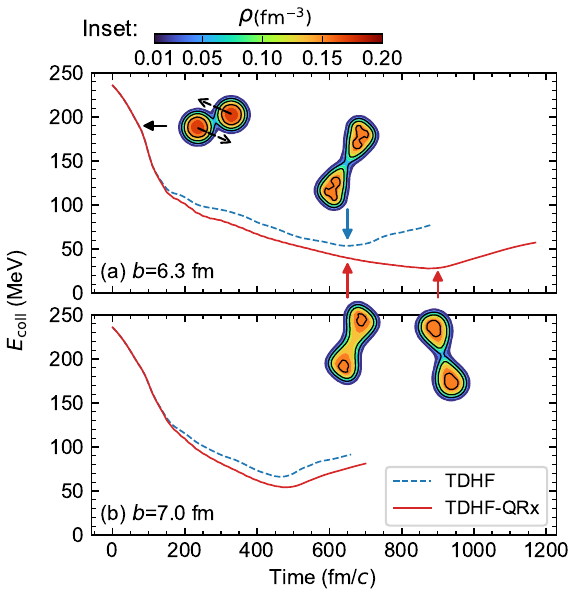} 
\caption{\label{fig:ecoll}The collective energies as a function of time for $^{60}\rm{Ni}+{^{60}\rm{Ni}}$ at $b=6.3$ and $7.0$ fm, $E_{\rm{c.m.}}=300~\rm{MeV}$.
The insets show the system density $\rho$ at indicating moment.
The dashed arrows in the left inset indicate the direction of motion of two nuclei.}
\end{figure}
\begin{figure}
\center
\includegraphics[width=7.0cm]{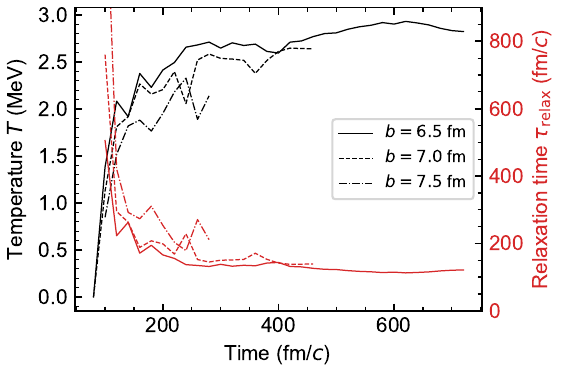} 
\caption{\label{fig:t_tau}The global temperature $T$ (black) and the relaxation time $\tau_{\rm relax}$ (red) as a function of time for $^{60}\rm{Ni}+{^{60}\rm{Ni}}$ at $E_{\rm{c.m.}}=300~\rm{MeV}$ with various impact parameters, calculated by TDHF-QRx.}
\end{figure}

The calculations are then performed on the same system at $E_{\rm{c.m.}}=300~\rm{MeV}$ ($\sim3.1V_{\rm B}$) with different impact parameters $b$.  
Figure~\ref{fig:ecoll} compares the time evolution of the collective energy $E_{\rm{coll}}$ at $b=6.3$ and $7.0$ fm, to illustrate the impact of two-body dissipations on the energy transfer process. 
The $E_{\rm{coll}}$ is calculated as
\begin{equation}
E_{\mathrm{coll}}=\frac{\hbar ^2}{2m}\int{\mathrm{d}^3r\frac{\boldsymbol{j}(\boldsymbol{r})^2}{\rho(\boldsymbol{r})}}.
\end{equation}
The two-body dissipations do not exert immediate influence when the nuclei first come into contact. 
Their impact on collective energy transformations remains slight during the intermediate stages of the reaction (e.g., $t = 150$-$650 \, \text{fm}/c$ in Fig.~\ref{fig:ecoll}(a)).
However, the two-body dissipations cause a longer contact time, and then a larger accumulated loss of $E_{\rm{coll}}$.
This different contact times suggest the potential influences of two-body dissipations on the equilibrium of quantities, like charge and mass, between the projectile and target nucleus \cite{PhysRevLett.124.212504}.
The impact on the nuclear shape at the scission point can also be observed in the insets of Fig.~\ref{fig:ecoll}, where TDHF-QRx indicates a more relaxed shape of the fragments.
For larger impact parameters, the effects of two-body dissipation on the reaction path become less significant, as shown in Fig.~\ref{fig:ecoll}(b).
Within the model framework, this is primarily attributed to the reduced overlap between the two nuclei, which leads to diminished momentum anisotropy and a lower temperature used in the $\tau_{\rm relax}$ calculation, as shown in Fig.~\ref{fig:t_tau}. 
However, since the local temperature in overlap region may remain higher in reality, the two-body dissipation effects could be underestimated by present model for such systems with strong localization.

\begin{figure}
\center
\includegraphics[width=7.5cm]{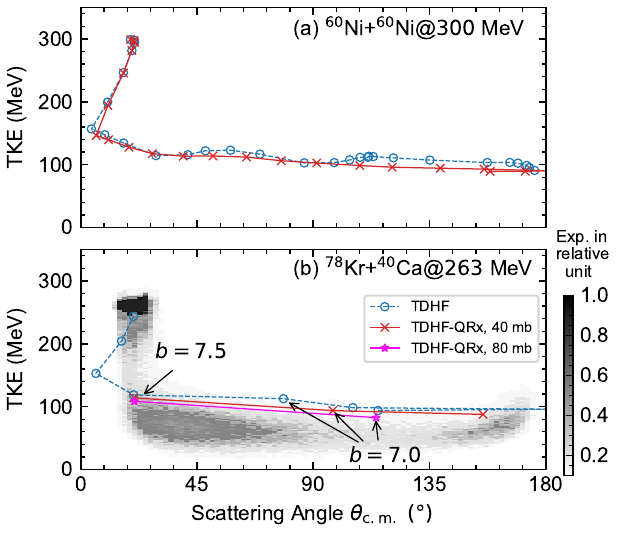} 
\caption{\label{fig:TKE}The fragments TKE-$\theta_{\rm c.m.}$ correlations for (a) $^{60}\rm{Ni}+{^{60}\rm{Ni}}$ at $E_{\rm{c.m.}}=300~\rm{MeV}$, and (b) $^{78}\rm{Kr}+{^{40}\rm{Ca}}$ at $E_{\rm{c.m.}}=263~\rm{MeV}$.
The cross-section values in legend indicate the corresponding $\sigma_{NN}$ in TDHF-QRx calculations.
The experimental data shown in background are taken from Ref.~\cite{Pirrone2019}.
The unit of impact parameter $b$ is fm.}
\end{figure}
The influences to the reaction process accumulate finally on the fragments.
Figure~\ref{fig:TKE}(a) illustrates the correlations between the total kinetic energy (TKE) and the scattering angle $\theta_{\rm c.m.}$ of fragments (the Wilczynski plot), usually used to exhibit the dissipations of reactions \cite{PhysRevLett.120.022501,PhysRevC.24.309}. 
Each point corresponds to one TDHF or TDHF-QRx simulation.
The interval of impact parameter between points, $\Delta b$, is 0.1 fm for $b < 7.5$ fm and 0.5 fm elsewhere.
At large impact parameters, the TKEs are close to the $E_{\rm{c.m.}}$, indicating that the reaction is elastic or quasielastic, without considerable conversion of collective energy into intrinsic energy. 
As impact parameter decreases, the reactions become inelastic. 
Due to the damping between the two nuclei, the scattering angle firstly approaches $0^{\circ}$ and then increases.
Notably, the impact on the TKE is relatively minor, whereas the increase in $\theta_{\rm c.m.}$ is more pronounced due to the extended contact time.
Besides, TDHF predicts an event clustering at $\theta_{\rm c.m.}\approx100^{\circ}$, which is not shown in TDHF-QRx calculations.

To compare with the experimental data, we also calculate the Wilczynski plot of $^{78}\rm{Kr}+{^{40}\rm{Ca}}$ at $E_{\rm{c.m.}}=263~\rm{MeV}$ ($\sim2.9V_{\rm B}$), shown in Fig.~\ref{fig:TKE}(b).
The $\Delta b$ between points is 0.5 fm.
For the inelastic cases, the TDHF has well agreement with the experimental TKE and the $\theta_{\rm c.m.}$, since the two-body dissipation effects are not significant for that cases. 
For the deep-inelastic cases, the TDHF calculations notably underestimate the energy dissipation as we expect about the limitation of TDHF.
Similar to the case of $^{60}\rm{Ni}$, the results of TDHF-QRx show slight enhancement of energy dissipation due to inclusion of two-body dissipation, however, still underestimate the actual dissipation.
Even though increasing $\sigma_{NN}$ can enhance energy dissipation, the underestimation is not attributed to a mismatch of $\sigma_{NN}$, as $\sigma_{NN}$ is closely linked to the critical angular momentum $L_{\rm fus}$ shown in Tab.~\ref{Table1}.
Larger $\sigma_{NN}$ leads to larger $L_{\rm fus}$, showing the effects of two-body dissipation in increasing the fusion cross-section, which is also found in Ref.~\cite{PhysRevC.93.034607}.
The experimental TKE for deep-inelastic events is $\sim60$ MeV, approximately being equal to the calculated Coulomb interaction energy at scission point. 
This indicates that the collective energy is almost entirely dissipated into intrinsic energy.
The deviation from experiment suggests that the current implementation of two-body dissipation, though capable of capturing momentum thermalization, is limited in accounting for all energy transfer channels in high-energy reactions. 
Further model refinement or inclusion of additional beyond-mean-field effects is required.

\begin{table}\center
\begin{tabular}{l c} 
 \hline
	& 		$L_{\rm fus}$ [$\hbar$]  	\\ 
 \hline
 Exp. \cite{Pirrone2019}		& 	117.0 			\\ 
 TDHF 	&	 106.2 		\\ 
 TDHF-QRx, 40 mb 	&	 117.2 		 \\ 
 TDHF-QRx, 80 mb 	&	 124.5 		\\ 
 TDHF-QRx, 100 mb 	&	 126.4 		\\ 
 \hline
\end{tabular}
\caption{The experimental and calculated critical fusion angular momentum $L_{\rm fus}$.
The cross-section values indicate the corresponding $\sigma_{NN}$ in TDHF-QRx calculations.
}
\label{Table1}
\end{table}

\begin{figure}
\center
\includegraphics[width=7.8cm]{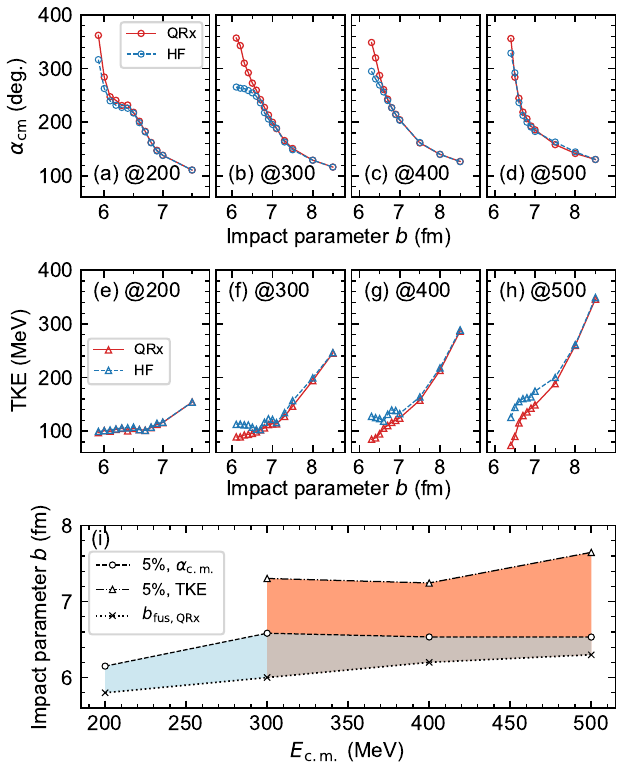} 
\caption{\label{fig:div} The comparison between TDHF and TDHF-QRx calculations of (a)-(d) rotation angle $\alpha_{\rm c.m.}$ and (e)-(h) TKE as a function of impact parameter $b$, for $^{60}\rm{Ni}+{^{60}\rm{Ni}}$ binary collisions at $E_{\rm{c.m.}}=$200, 300, 400, and 500 MeV.
(i) The reaction condition range where the impact of two-body dissipation to $\alpha_{\rm c.m.}$ (blue) and TKE (orange) exceeds $5\%$, for the $^{60}\rm{Ni}+{^{60}\rm{Ni}}$ binary collision.}
\end{figure}
To estimate the region where the momentum-space thermalization of two-body dissipation affects mean-field dynamics,
we extend the calculation of $^{60}\rm{Ni}+{^{60}\rm{Ni}}$ binary reaction to energies 200, 400, and 500 MeV.
Figure~\ref{fig:div}(a) and (b) illustrate the comparison of calculated rotational angle $\alpha_{\rm c.m.}$ and the TKE between TDHF and TDHF-QRx. 
The $\alpha_{\rm c.m.}$ is defined as the rotation angle of the vector formed by the mass centers of the two nuclei from the initial to the final state. Its relation with the scattering angle $\theta_{\rm c.m.}$ is 
\begin{equation}
\theta_{\rm c.m.}=\left | \pi-\theta_{\rm in}-\theta_{\rm out}-\alpha_{\rm c.m.} \right |,
\end{equation} 
where $\theta_{\rm in}$ and $\theta_{\rm out}$ denote the incoming and outgoing Rutherford angles, respectively. 
$\alpha_{\rm{c.m.}}$ is adjusted to the range $[0, 2\pi)$ by subtracting $2n\pi$ if it exceeds $2\pi$.
The rotational angle $\alpha_{\rm c.m.}$ is chosen due to its more intuitive correlation with the contact time compared to $\theta_{\rm c.m.}$. 
At large $b$, deviations between TDHF and TDHF-QRx remain insignificant across all energy cases, suggesting that the impact of the momentum thermalization is limited in grazing collisions even at higher energies, due to the limited overlap between the two nuclei. 
However, as discussed in Fig.~\ref{fig:t_tau}, the possibility of an underestimation of two-body dissipation cannot be excluded.
As $b$ decreases, the thermalization effects induced by two-body dissipation become more pronounced, resulting in larger deviations, except at the lowest energy case (200 MeV), where the excitation energy is insufficient to induce significant two-body dissipation. 
We define the relative deviation between TDHF and TDHF-QRx calculations as
\begin{equation}
\Delta X _{\mathrm{rel}}=\frac{\left| X _{\mathrm{HF}}-X_{\mathrm{QRx}} \right|}{X _{\mathrm{HF}}}. 
\end{equation} 
By interpolation, the critical $b$ corresponding to a $5\%$ relative deviation can be determined, represented as dashed ($\alpha_{\rm c.m.}$) and dash-dotted (TKE) line in Fig.~\ref{fig:div}(i).
It enables identification of the reaction condition range where two-body dissipation should be carefully considered (relative deviation exceeds $5\%$), as shown by the color-filled regions.  
Reactions below the critical fusion $b$ (dotted line) fall into the fusion regime, where no fragments are present to determine relative deviation percentages, but the two-body dissipation also contributes to the thermalization of the pre-equilibrium stage.

The results offer insight into the possible influence of two-body dissipation effects on the mean-field approach.
They confirm the validity of mean-field approximation near the Coulomb barrier and, more importantly, delineate the probable boundary. 
The $b$ value of this boundary seems to exhibit an insensitivity to $E_{\rm{c.m.}}$ in high energy cases.
As nuclei are finite quantum systems, the role of two-body dissipation can vary between reaction systems, potentially altering Fig.~\ref{fig:div} for different cases. 
Exploring the systematic trends across different reaction systems presents an exciting direction for future work.

It is worth discussing the limitations inherent in the current modeling framework when describing strongly localized systems.
The calculation results should be interpreted in light of certain limitations of the model.
The linear form of NN collisions in RTA is under the assumption of near-equilibrium conditions, and a non-local temperature is employed in the relaxation time.
Both of these factors may lead to an underestimation of two-body dissipation when the system is far from equilibrium.
Additionally, the local character of NN collisions is not fully respected by replacing the local kinetic energy with the total energy as the constrained quantity in the equilibrium-state calculation, which may result in spurious thermal relaxation outside the NN collision region.

\section{Summary}
\label{summary}

Microscopic mean-field dynamics is well-suited for low-energy nuclear reactions but its validity in higher-energy reactions, especially concerning two-body dissipation effects, remains uncertain.
In this letter, we develop the TDHF-QRx approach by incorporating the quantum relaxation-time approximation to address the nucleon-nucleon collision term, enabling a direct evaluation of momentum-space thermalization effects from two-body dissipation on the mean-field dynamics.

The primary impact of two-body dissipation is the extended contact time between colliding nuclei, which could enhance the equilibration of physical quantities during the reaction and alters final fragment properties, such as scattering angles. 
Using the $^{60}\rm{Ni}+{^{60}\rm{Ni}}$ binary reaction as an example, we identified the conditions where two-body dissipation becomes significant, offering valuable insights into the boundary of mean-field approximation in microscopic reaction dynamics.

Although the present model is capable of capturing the momentum thermalization from two-body dissipation, further refinement is required to address the limitations introduced by its assumptions and to better capture the complex dissipation dynamics in high-energy nuclear reactions, particularly for the systems with strong localized characters.

\section*{Acknowledgements}
We thank Xiang Jiang, Zhen Zhang, Yinu Zhang, Zepeng Gao, and Zehong Liao for valuable discussions.
This work is supported by the National Natural Science Foundation of China under Grant No. 12475136.



\bibliographystyle{elsarticle-num} 
\bibliography{biblio}





\end{document}